\documentclass[dvips]{article}
\usepackage{supertabular,lscape,epsfig}
\usepackage{amssymb}
\usepackage{amsmath}
\DeclareSymbolFont{ppa}{OT1}{ppl}{m}{it}
\DeclareMathSymbol{\vv}{\mathalpha}{ppa}{'166}

\begin{document}

\newcommand{\dd}{\,{\rm d}}
\newcommand{\ie}{{\it i.e.},\,}
\newcommand{\etal}{{\it et al.\ }}
\newcommand{\eg}{{\it e.g.},\,}
\newcommand{\cf}{{\it cf.\ }}
\newcommand{\vs}{{\it vs.\ }}
\newcommand{\zdot}{\makebox[0pt][l]{.}}
\newcommand{\up}[1]{\ifmmode^{\rm #1}\else$^{\rm #1}$\fi}
\newcommand{\dn}[1]{\ifmmode_{\rm #1}\else$_{\rm #1}$\fi}
\newcommand{\upd}{\up{d}}
\newcommand{\uph}{\up{h}}
\newcommand{\upm}{\up{m}}  
\newcommand{\ups}{\up{s}}
\newcommand{\arcd}{\ifmmode^{\circ}\else$^{\circ}$\fi}
\newcommand{\arcm}{\ifmmode{'}\else$'$\fi}
\newcommand{\arcs}{\ifmmode{''}\else$''$\fi}
\newcommand{\MS}{{\rm M}\ifmmode_{\odot}\else$_{\odot}$\fi}
\newcommand{\RS}{{\rm R}\ifmmode_{\odot}\else$_{\odot}$\fi}
\newcommand{\LS}{{\rm L}\ifmmode_{\odot}\else$_{\odot}$\fi}

\newcommand{\Abstract}[2]{{\footnotesize\begin{center}ABSTRACT\end{center}
\vspace{1mm}\par#1\par   
\noindent
{~}{\it #2}}}

\newcommand{\TabCap}[2]{\begin{center}\parbox[t]{#1}{\begin{center}
  \small {\spaceskip 2pt plus 1pt minus 1pt T a b l e}
  \refstepcounter{table}\thetable \\[2mm]
  \footnotesize #2 \end{center}}\end{center}}

\newcommand{\TableSep}[2]{\begin{table}[p]\vspace{#1}
\TabCap{#2}\end{table}}

\newcommand{\FigCap}[1]{\footnotesize\par\noindent Fig.\  %
  \refstepcounter{figure}\thefigure. #1\par}

\newcommand{\TableFont}{\footnotesize}
\newcommand{\TableFontIt}{\ttit}
\newcommand{\SetTableFont}[1]{\renewcommand{\TableFont}{#1}}

\newcommand{\MakeTable}[4]{\begin{table}[htb]\TabCap{#2}{#3}
  \begin{center} \TableFont \begin{tabular}{#1} #4
  \end{tabular}\end{center}\end{table}}

\newcommand{\MakeTableSep}[4]{\begin{table}[p]\TabCap{#2}{#3}
  \begin{center} \TableFont \begin{tabular}{#1} #4
  \end{tabular}\end{center}\end{table}}

\newenvironment{references}%
{
\footnotesize \frenchspacing
\renewcommand{\thesection}{}
\renewcommand{\in}{{\rm in }}
\renewcommand{\AA}{Astron.\ Astrophys.}
\newcommand{\AAS}{Astron.~Astrophys.~Suppl.~Ser.}
\newcommand{\ApJ}{Astrophys.\ J.}
\newcommand{\ApJS}{Astrophys.\ J.~Suppl.~Ser.}
\newcommand{\ApJL}{Astrophys.\ J.~Letters}
\newcommand{\AJ}{Astron.\ J.}
\newcommand{\IBVS}{IBVS}
\newcommand{\PASP}{P.A.S.P.}
\newcommand{\Acta}{Acta Astron.}
\newcommand{\MNRAS}{MNRAS}
\renewcommand{\and}{{\rm and }}
\section{{\rm REFERENCES}}
\sloppy \hyphenpenalty10000
\begin{list}{}{\leftmargin1cm\listparindent-1cm
\itemindent\listparindent\parsep0pt\itemsep0pt}}%
{\end{list}\vspace{2mm}}
 
\def\TYLDA{~}
\newlength{\DW}
\settowidth{\DW}{0}
\newcommand{\dw}{\hspace{\DW}}

\newcommand{\refitem}[5]{\item[]{#1} #2%
\def\REFARG{#3}\ifx\REFARG\TYLDA\else, {\it#3}\fi
\def\REFARG{#4}\ifx\REFARG\TYLDA\else, {\bf#4}\fi
\def\REFARG{#5}\ifx\REFARG\TYLDA\else, {#5}\fi.}

\newcommand{\Section}[1]{\section{#1}}
\newcommand{\Subsection}[1]{\subsection{#1}}
\newcommand{\Acknow}[1]{\par\vspace{5mm}{\bf Acknowledgements.} #1}
\pagestyle{myheadings}

\newfont{\bb}{ptmbi8t at 12pt}
\newcommand{\xrule}{\rule{0pt}{2.5ex}}  
\newcommand{\xxrule}{\rule[-1.8ex]{0pt}{4.5ex}}  
\def\thefootnote{\fnsymbol{footnote}}
\begin{center}

{\Large\bf 
Eclipsing binaries in the open cluster NGC 2243 - I. Photometry\footnote
{This paper is based in part on data obtained at the 
South African Astronomical Observatory.}
}
\vskip1cm
{\bf
J.~~~K~a~l~u~z~n~y$^1$,
~~W.~~K~r~z~e~m~i~n~s~k~i$^2$ ,
~~I.~B.~~T~h~o~m~p~s~o~n$^3$,
 and~~ G.~~S~t~a~c~h~o~w~s~k~i$^{1}$\\}
\vskip3mm
{
  $^1$Nicolaus Copernicus Astronomical Center,
     ul. Bartycka 18, 00-716 Warsaw, Poland\\
     e-mail: (jka,gss@camk.edu.pl)\\
  $^2$Las Campanas Observatory, Casilla 601, La Serena, Chile,\\
     e-mail: (wojtek@lco.cl)}\\
  $^3$The Observatories of the Carnegie Institution of Washington,
     813 Santa Barbara Street, Pasadena, CA 91101, USA\\
     e-mail: (ian@ociw.edu)\\
\end{center}

\Abstract{We obtained $BV$ time series photometry for 12 variable
stars from the field of the old open cluster NGC~2243. 
The sample includes 3 newly identified detached/semi-detached binaries. 
There are now four detached eclipsing binaries which are likely 
members of the cluster. Determination of the absolute parameters 
of the components would provide a valuable check on evolutionary
models of low-mass stars.  
An accurate ephemeris and orbital period analysis are
presented for the previously-known detached 
binary NV~CMa. We also provide ephemerides for 7 other periodic variables.
We show that 3 contact binaries
are likely members of the cluster. 
}
{Stars:  binaries: eclipsing, binaries -- stars: individual: NV CMa  -- open
clusters and associations: individual: NGC~2243}
\Section{Introduction}

Detached eclipsing binaries (DEBs)  at the moment provide the most accurate 
determinations of the absolute stellar parameters of mass and radius (Andersen 1991, 1998).
Although recent developments in high-precision astrometry,  combined with 
radial velocity measurements, hold great promise in this 
respect (eg. Muterspaugh et al. 2005), DEBs can be observed at
much larger distances than visual binaries and, moreover, they allow a more 
direct determination of stellar radii and temperatures.
Systems of early spectral type can be used as standard candles as far as 
in the Local Group Galaxies (Paczy\'nski 1997; Clausen 2004; Ribas et al. 2005).
DEBs located in open and globular clusters provide a particularly interesting
check on stellar evolution theory (Paczy\'nski 1997).     

NGC~2243 is an intermediate age open cluster located in the direction
of Galactic anticenter ($l=239^{\circ}.5$, $b=-18^{\circ}.0$) in a
region of low reddening. An early photometric study of the cluster by
Hawarden (1975) was followed by CCD-based studies by Bonifazi (1990)
and Bergbusch et al. (1991), and more recently by Anthony-Twarog et al.
(2005).  The last authors used $uvbyCaH{\beta}$ photometry to derive
reddening $E(B-V)=0.055\pm 0.004$, metallicity ${\rm [Fe/H]}=-0.57\pm
0.03$, age of $3.8\pm 0.2$~Gyr, and an apparent distance modulus
$(m-M)_{V}=13.15\pm 0.1$.  Spectroscopic studies conducted by Gratton
\& Contarini (1994; only two giants observed) and  Friel et al. (2002)
gave ${\rm [Fe/H]}=-0.48\pm 0.15$ and ${\rm [Fe/H]}=-0.49\pm 0.05$,
respectively.  The CCD-based color magnitude diagrams of NGC~2243 (see
references above) exhibit a well marked sequence of `photometric
binaries', formed by stars located about 0.7 mag above the main
sequence defined by the `single' stars. This indicates that the cluster
possesses a rich population of binary stars with mass ratio close to
unity.

The field of NGC~2243 was surveyed for variable stars by Kaluzny et al.
(1996; hereafter KKM).  The sample of 6 detected objects included 3
DEBs, 2 contact binaries and a background RR Lyr star. Recently,
Anthony-Twarog et al. (2005) identified 4 new candidate variables in
the cluster field.

The present study resulted from an effort aimed mainly at obtaining a
good quality light curve for the eclipsing binary V1=NV~CMa detected by
KKM. The system is located near the main-sequence turnoff on the
cluster color-magnitude diagram. Its light curve exhibits two eclipses
of similar depth with $\delta V\approx 0.6$~mag. The flatness of the
light curve outside of eclipses together with an orbital period of
$P=1.19$~days indicates that the binary has a detached configuration
and hence deserves more detailed study.  As a first step toward this
goal we present $BV$ light curves of V1=NV~CMa. We  also report some
results for other objects, including the detection of new variables in
the central field of the cluster.

\Section{Observations and Data Reduction}

New observations aimed primarily at obtaining complete $BV$ light
curves of NV~CMa were obtained with the 1.0-m Swope telescope at Las
Campanas Observatory.  The telescope was used at the f/13.5 focus
without a Gascoigne corrector lens which is employed in its normal f/7
configuration.  The atypical f/13.5 configuration allowed better
sampling of the point spread function (PSF) at a cost of a reduced
field of view.  It was considered important to obtain well sampled
images as NV~CMa possesses a close visual companion located at an
angular distance of about 1.6 arcsec. It would be difficult to obtain
accurate time series photometry of the variable if the stellar images
were poorly sampled.  The first data set was collected on 6 nights
between 1995 December 29 and 1996 January 4 (run\#1).  We used the
Tektronix $1024\times1024$ TEK1 detector giving a scale 0.36
arcsec/pix.  Exposure times ranged from 90s to 120s for the $V$ filter
and from 90 s to 240 s for the $B$ filter.  A total of 55 $B$-band and
689 $V$-band images were collected.
The second observing run occurred during the period 1997 January 6-10
(run\#2). The Tektronix $2048\times2048$ TEK5 detector subrastered to
$1200\times 1200$ pixels was used. The scale was 0.33 arcsec/pixel.
The exposure time was set to 200s for the $B$ band and to 120s for the
$V$ band.  A total of 160 $B$-band and 225 $V$-band
images were collected.
During runs \#1 and \#2 the cluster was monitored  for
a total of about 57 hours. 

Some additional data aimed at a photometric
calibration of the cluster field were obtained on 2 nights in 1999 November.
The $2048\times4096$ SITe3 detector subrastered to $2048\times3150$ 
pixels was used
at the f/7 focus of the Swope telescope (run \#3). This configuration 
resulted in images with a scale of 0.435 arcsec/pixel. 

The last set of observations was 
collected with the 1.0-m telescope at SAAO observatory (run\#4). 
These observations were directed toward a photometric calibration of the 
cluster field. An additional goal was to improve the ephemeris
for NV~CMa by determining the moments of two minima of light.
The detector was a STE4 $1024\times 1024$ camera giving a scale of 0.31 
arcsec/pixel. During that run the cluster was observed for a total of 
9 hours on 5 nights spanning the period 2005 October 18-25. 

The raw data were pre-processed with the IRAF-CCDPROC
package.\footnote{IRAF is distributed by the National Optical Astronomy
Observatories, which are operated by the Association of Universities
for Research in Astronomy, Inc., under cooperative agreement with the
NSF.} In particular frames obtained during run \#3 were corrected for
the known non-linearity of the SITe3 camera.  Time series photometry
was extracted using the ISIS-2.1 image subtraction package (Alard \&
Lupton 1998; Alard 2000). Our procedure followed closely that described
in some detail by Mochejska et al.  (2002). Instrumental magnitude zero
points for the ISIS differential light curves were determined from the
template images using the DAOPHOT/ALLSTAR package (Stetson 1987).
Aperture corrections were measured with the DAOGROW program (Stetson
1990). Instrumental magnitudes were transformed to the standard $BV$
system using a large assembly of local standards established during run
\#3 (see below).


\Section{Photometric Calibration}

We first attempted to obtain a photometric calibration for the
monitored field during run \#2. Unfortunately, later analysis showed
that all nights during that run were only marginally photometric.  To
resolve this, we observed NGC~2243 along with several Landolt (1992)
fields on two photometric nights, 1999 November 16 and 18, during run
\#3.  Transformation coefficients calculated for these two nights are
consistent with each other.  Further standards were observed on the
night of November 16. Specifically, on that night 7 $BV$ observations
of 3 standard fields (T Phe, RU 149 and RU 152) were collected. There
were a total of 41 standard stars in these fields with photometry
available from Stetson\footnote{The current version of the catalog is
available at http://cadcwww.hia.nrc.ca/astrocat. Throughout this paper
we use Stetson's photometry dated at May 2 2005.} (2000).  The
following relations between instrumental and standard magnitudes were
obtained:
\begin{eqnarray}
v=V-0.010(3)\times (B-V)+0.15\times (X-1.25)+ 2.967(2), \\
b-v=0.965(3)\times (B-V)+0.12\times (X-1.25)  +0.240(3),\\ 
b=B-0.049(3)\times (B-V) +0.27\times(X-1.25) + 3.208(2).
\end{eqnarray} 
The standards were observed over small range of air-mass, $1.06<X<1.20$.
Therefore, no attempt has been made to derive extinction coefficients. 
Instead average values of extinction at Las Campanas were adopted.
This has little effect on the calibration of NGC~2243 field as the
cluster was observed at an air mass of $X=1.12$, which falls within 
the range of air-masses covered by the standards. 
In Fig. 1 we show residuals between the standard and recovered 
magnitudes and colors for Stetson standards from Landolt's fields.
\begin{figure}[htb]
\centerline{\includegraphics[height=80mm,width=120mm]{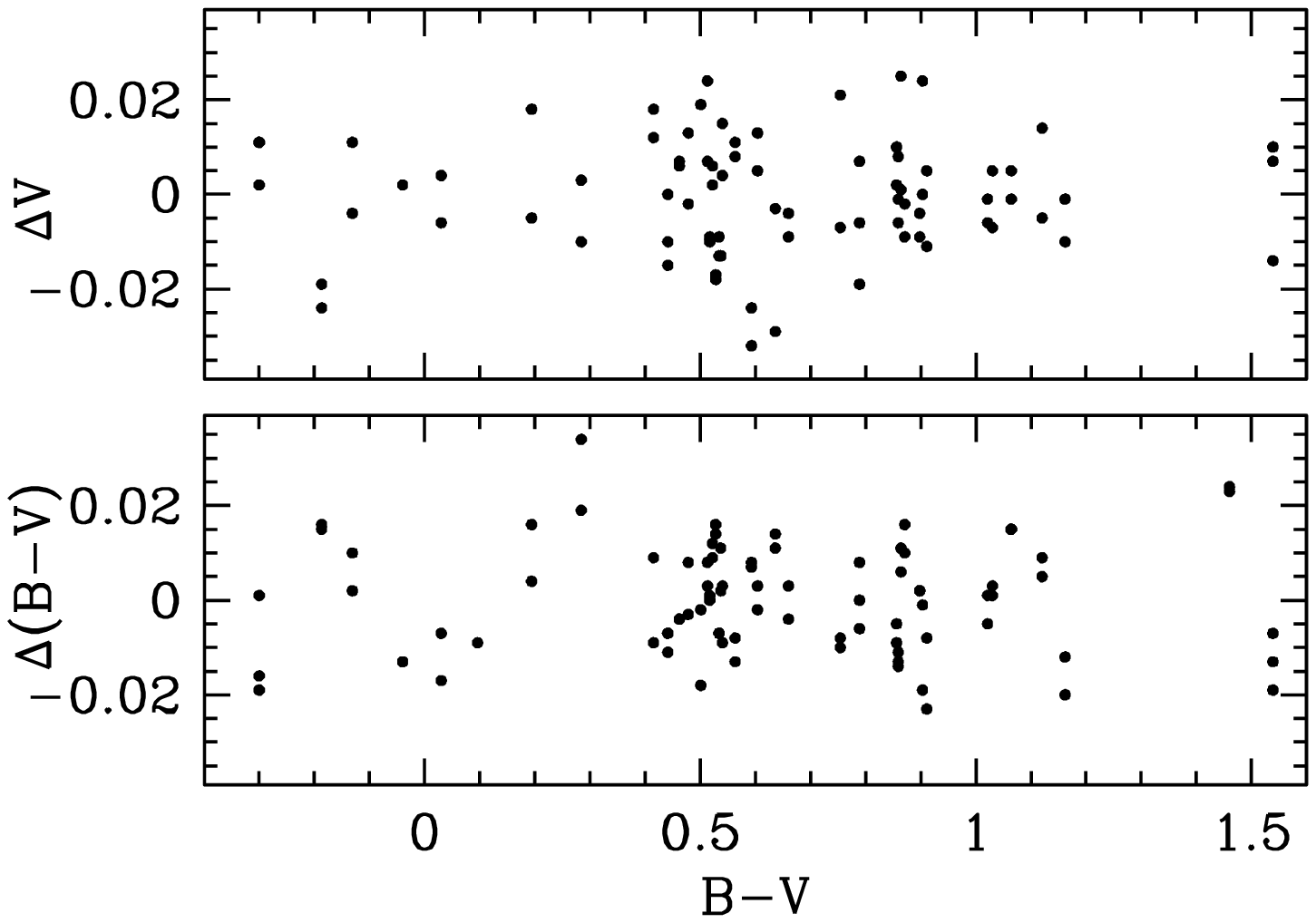}}
\FigCap{
Plot of the color and magnitude residuals for the standard stars 
observed on the night of 1999 November 16.
}
\end{figure}
Aperture photometry was extracted for standards with the
DAOPHOT and DAOGROW programs  (Stetson 1987, 1990). As for the cluster,
field extraction of profile photometry with DAOPHOT/ALLSTAR was followed 
by determination of aperture corrections for frames with 
all stars subtracted
except those used for determination of the PSF.  Specifically,
aperture corrections were determined for a pair of NGC~2243 frames:
$V$--300s and $B$--420s. Photometry based on these frames was supplemented 
by measurements extracted from two shorter exposures: $V$--60s and $B$--90s.
The one sigma uncertainty of the zero points of our photometry is about
0.015 mag for both bands (this includes errors of transformation 
coefficients and errors of aperture corrections).
Cluster stars with $BV$ photometry based on the data from run \#3
were subsequently used as local standards. In particular 
instrumental photometry from runs \#1 and \#2 was tied to these local 
standards.
The observations obtained on 1999 November 18 resulted in $BV$
photometry of NGC~2243 with an average difference  in the zero points
with respect
to the November 16 data of $\Delta V=0.001$ and $\Delta (B-V)=0.002$.

As shown below, there is an unexpectedly large offset of the zero 
point of our $V$ photometry relative to Stetson (2000). To clarify
this issue an effort was made to obtain yet another independently
calibrated set of $BV$ photometry of NGC~2243. During run \#4 
on the night of 2005 October 20, we observed the cluster along with
five Landolt fields containing a total of 21 $BV$ standards. 
The air-masses covered by the observations of the standards spanned the range 
1.12--1.76. Each of the Landolt fields was observed several times and 
we obtained a total of 66 measurements of standards for each of the filters.
The following relations between instrumental and standard magnitudes were
obtained:
\begin{eqnarray}
v=V-0.020(3)\times (B-V)+0.140(7)\times (X-1.25)+ 2.572(2), \\
b-v=1.073(6)\times (B-V)+0.125(16)\times (X-1.25) + 0.484(5).
\end{eqnarray}
The cluster was observed at an air-mass of $X=1.01$. As shown 
below, $BV$ photometry obtained at SAAO remains in good
agreement with results obtained at LCO.

\subsection{Comparison with Previous Photometry}

Three sets of CCD based $BV$ photometry of NGC~2243 were published so far.
Bonifazi et al. (1990) calibrated their data using local standards
from the cluster field established by van den Bergh (1977). 
Bergbusch et al. (1991) tied their photometry to standards from Landolt (1983)
and Graham (1982). Photometry for 14 stars included in 
Stetson (2000) data base is tied to Landolt (1992) standards.
In addition, Anthony-Twarog et al. (2005) list $V$ magnitudes for 1267 stars
observed on the $uvby$ system. 
Table 1 summarizes the comparison of our photometry from run \#3 with 
the data sets listed above. Also included is a comparison with photometry
obtained by us at SAAO (run\#4). The mean residuals, in the sense "run \#3" 
minus references, were calculated for stars brighter than $V=19.5$.
All stars with residuals above or below 0.05 mag from the mean were 
excluded from an analysis using an iterative procedure. 
No significant systematic dependence of residuals on color could be
noted in all but two  cases: a systematic trend is seen for $\Delta (B-V)$
for Bonifazi et al. (1990); also,  some systematic
deviation is present for $\Delta V$ in case of the reddest stars  
from the sample of Anthony-Twarog et al. (2005).
A plot illustrating these two cases is shown  in Fig. 2.
\begin{figure}[htb]
\centerline{\includegraphics[height=45mm,width=120mm]{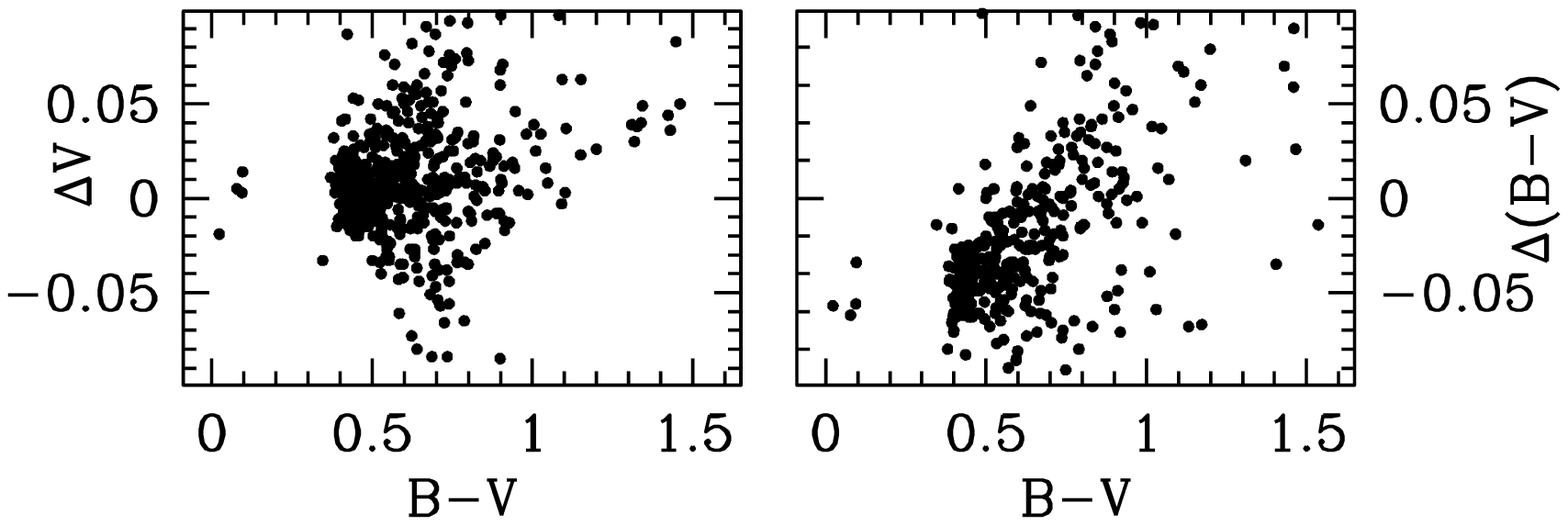}}
\FigCap{
Left -- differences between our magnitudes and those 
obtained by Antony-Twarog et al. (2005). Right -- differences between our
colors and those obtained by Bonifazi et al. (1990). 
}
\end{figure}
As for the zero point of the $V$ magnitude, there is generally good 
agreement between  all the sources of photometry considered 
except that of Stetson (2000).   
Both our data sets imply slightly bluer colors than 
the remaining 3 surveys. However, note that for Bonifazi et al. (1990)
the $\Delta (B-V)$ is correlated with the color.
\begin{table}
\centering
\caption{\small Comparison with other $BV$ surveys of NGC~2243 }
{\small
\begin{tabular}{llrlr}
\hline
Source & $\Delta V$ & N$_{\rm V}$ & $\Delta (B-V)$  &  N$_{\rm (B-V)}$ \\
\hline
Bonifazi et al. (1990) &  -0.007(17)      & 291    & -0.035(21) & 287\\
Bergbusch et al. (1991) &   0.010(16)      & 347    & -0.040(20) & 343\\
Stetson (2000)         &  -0.083(9)       & 11     & -0.024(15) & 11\\
Anthony-Twarog et al. (2005) & 0.007(19) & 481    &       &     \\
run \#4                & 0.023(15)       & 244    & -0.002(18) & 236 \\
\hline
\end{tabular}}
\end{table}
\Section{Variables}

Times series photometry based on the data from runs \#1, \#2 and \#4
was analyzed with the TATRY program kindly provided by Alex Schwarzenberg-Czerny.
This program is suitable for the detection of various types of variable stars 
and uses in one of its modes the AoV algorithm (Schwarzenberg-Czerny 1996).
A total of 12 variables were identified. Their equatorial coordinates
tied to the astrometric frame of the USNO-B catalog are listed in Table 2.
Finding charts are presented in Fig. 3.
\begin{figure}[htb]
\centerline{\includegraphics[height=90mm,width=120mm]{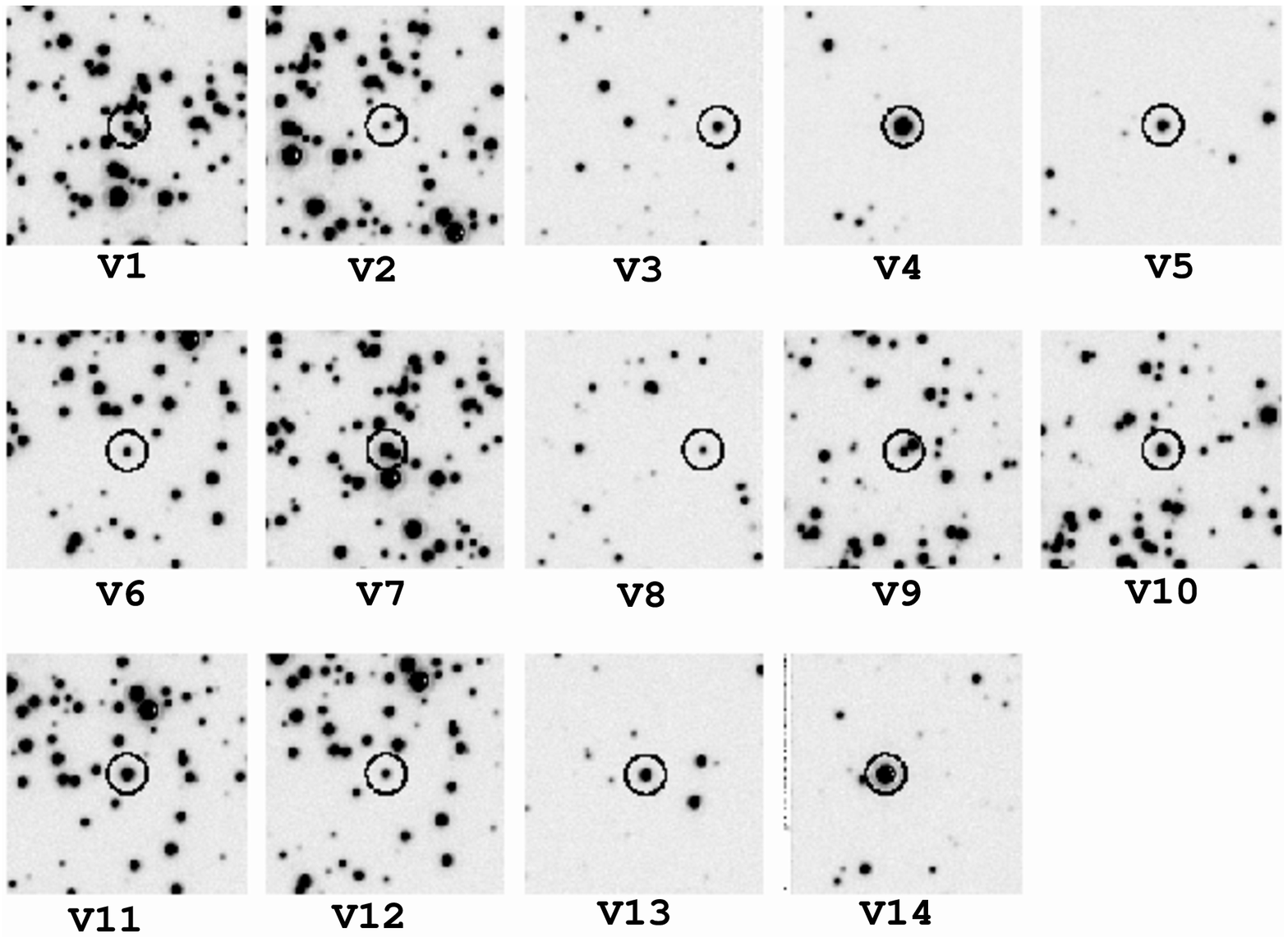}}
\FigCap{
Finder charts for NGC~2243 variables.
Each chart is 60 arcsec on a side. North is up, and east to the left. 
}
\end{figure}
Variables V1-6 were already listed in KKM. All but one of the  remaining 
objects are new identifications. Star V7  was noted earlier as
a candidate variable by Anthony-Twarog et al. (2005; their  candidate \#1746)
\footnote{Candidate variables \#1419 and \#1463 from Antony-Twarog et al. 
do not show any evidence for variability in our data. Candidate \#2445 
was not observed by us.}. 
We did not obtain any new time series photometry for eclipsing 
variables V4 and V5. These were observed out of eclipse during 
run \#3; we used these data for determination  of $V_{\max}$ 
and $(B-V)_{\rm max}$.\\ 
Table 3 lists ephemerides for objects whose periods could be 
unambiguously determined based on the available data.
Phased light curves for these stars are shown in Fig. 4.
%
\begin{table}
\centering
\caption{\small Coordinates and photometric parameters of 
variables from NGC~2243 field.}
{\small
\begin{tabular}{lccllll}
\hline
Name	 &RA$_{\rm 2000}$ &Dec$_{\rm 2000}$& $V_{\rm MAX}$& $\Delta V$& $(B-V)_{\rm MAX}$&Type\\
\hline
V1=NV~CMa &6:29:35.49& -31:16:53.6&  16.40& 0.61&  0.44& Ecl\\
V2=NU~CMa &6:29:33.81& -31:17:03.4&  17.86& 0.27&  0.66& Ecl-EW\\
V3=NW~CMa &6:29:44.92& -31:18:19.4&  16.69& 0.25&  0.43& Ecl-EW\\
V4=NS~CMa &6:29:09.59& -31:15:34.0&  14.11& 0.7:&  0.35& Ecl\\
V5=NX~CMa &6:29:56.07& -31:20:19.1&  16.34& 0.25:& 0.51& Ecl\\
V6=NT~CMa &6:29:33.58& -31:17:58.4&  16.89& 0.93&  0.16& RRab\\
V7	  &6:29:35.74& -31:17:04.3&  15.15& 1.05&  0.32& Ecl\\
V8   &6:29:45.33& -31:17:19.2&  18.39& 0.35&  0.61& Ecl\\
V9   &6:29:34.30& -31:16:18.1&  17.17& 0.16&  0.64& Ecl\\
V10   &6:29:33.44& -31:16:24.1&  15.88& 0.03&  0.45& $\delta$~Sct/Ecl-EW?\\
V11   &6:29:32.79& -31:17:46.1&  15.59& 0.04&  0.74& ?\\
V12  &6:29:33.01& -31:17:53.6&  17.38& 0.05&  0.57& Ecl-EW \\
V13  &6:29:27.11& -31:18:28.8&  16.07& 0.02&  0.43& $\gamma$~Dor?\\
V14  &6:29:41.09& -31:19:06.2&  13.75& $>0.09$& 1.03& ?\\
\hline
\end{tabular}}
\end{table}
%
\begin{table}
\centering
\caption{\small Ephemerides for periodic variables from NGC~2243 
field.}
{\small
\begin{tabular}{lll}
\hline
Name	 &Period[d] &       T0      \\
         &          &   HJD 2400000+\\
\hline
V1=NV~CMa &1.18851590(2) & 48663.70748(5) \\
V2=NU~CMa &0.2853011(1)  & 48224.8492(1) \\
V3=NW~CMa &0.3564557(1)  & 48225.0181(1) \\
V6=NT~UMa &0.5865921(2)  & 48225.0471(2) \\
V7	  &1.382703(2)   & 48225.7332(15) \\
V10   &0.12785(4)    & 50081.6711(1) \\
V12  &0.28598(11)   & 50081.6456(20) \\
V13  &0.76793(2)    & 50081.6077(15) \\
\hline
\end{tabular}}
\end{table}
%
\begin{figure}[htb]
\centerline{\includegraphics[height=130mm,width=120mm]{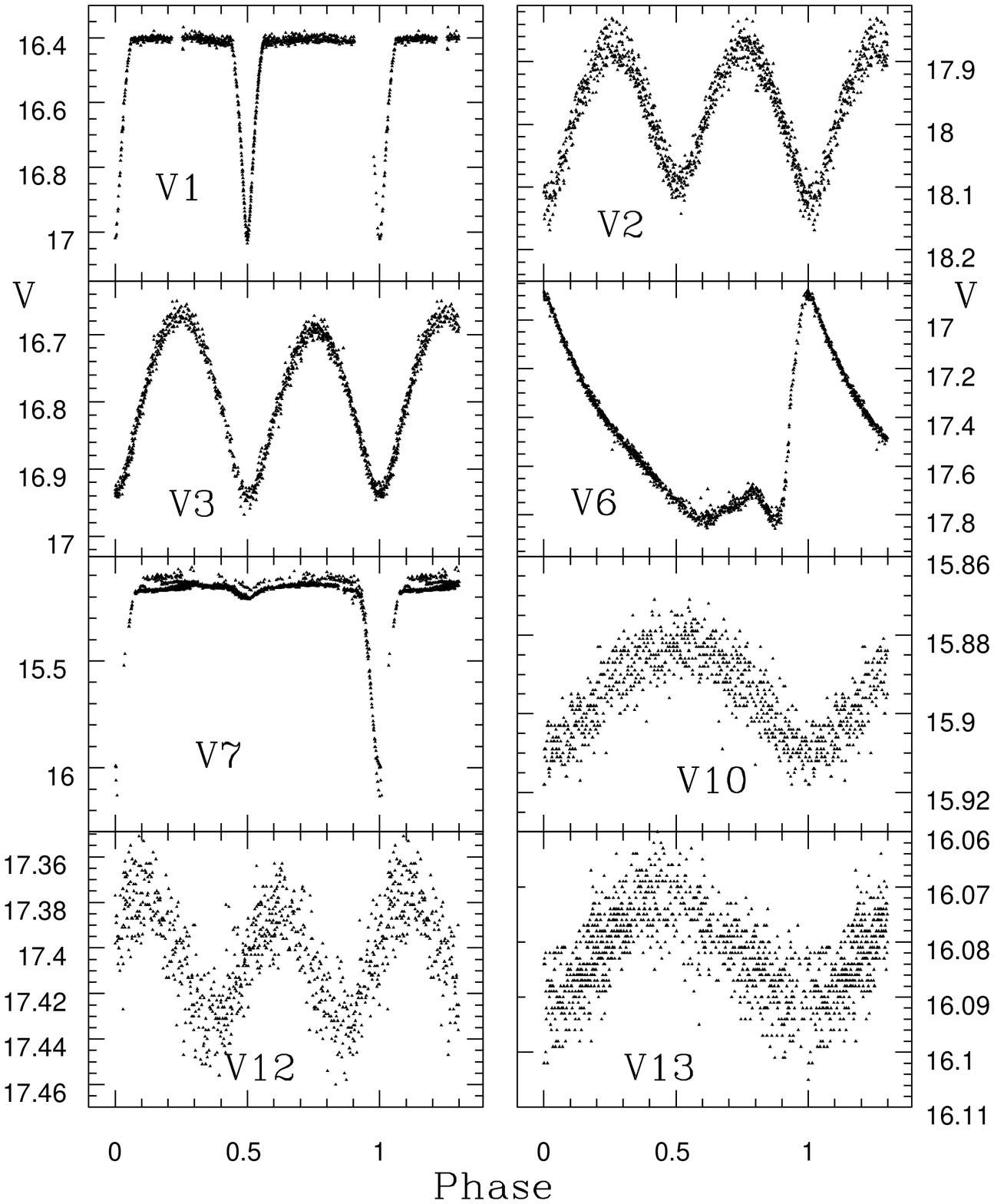}}
\FigCap{
Phased $V$ light curves of NGC~2243 variables. 
}
\end{figure}
Two eclipse events were observed for variable V8. The first 
was centered  at $HJD\approx 2448286.70$ while the second occurred
at $HJD\approx 2450457.835$. The light curve, including photometry from
all observing seasons,
can be phased with $P=6.2568$~d. However,  some longer periods
are also allowed by our data. The second of the observed eclipses of V8 is 
shown in Fig. 5
Two primary ($HJD\approx 2449342.56$,  $HJD\approx 2450456.57$) 
and one secondary ($HJD\approx 2450086.76$) eclipses were detected in 
the light curves of variable V9. Two of these eclipses are 
shown in Fig. 5. All observations of V9 
can be phased
with a period of $P=3.257344$. However the adoption of such a  period
would imply noticeable eccentricity
of the binary orbit. This is a rather unlikely configuration 
for a short-period main-sequence binary belonging to a cluster with an age of 
about 4 Gyr. Our spectroscopic
data (Kaluzny et al. 2006, in preparation) indicate that V9 is  indeed a radial
velocity member of NGC~2243, hence the listed period for V9
is most likely spurious. 

Variable V12 has been classified as an eclipsing contact binary of W~UMa
type; its color is too red for a $\delta $~Sct variable. 
The situation is more complicated in case of V10. With an unreddened color of
$(B-V)\approx 0.40$ it can be either a
$\delta $~Sct star with $P\approx 0.128$~d or a low-amplitude contact  
binary with twice that period. The available photometric data do not allow
us to distinguish between these two possibilities.\
The variable V13 has been preliminarily classified as a $\gamma$~Dor type
pulsator
based on its period and the shape of the light curve. \\
Time domain light curves of the unclassified variables V11 and V14
are presented in Fig. 6. In case of V11 useful data were collected 
only during  run \#1. The photometry from run \#2 is 
affected by some instrumental effects,  
while that from KKM is rather noisy
(note the small amplitude of variability). In the case of variable V14 useful
photometry was obtained  only during run \#2. 
During all other runs this relatively bright star was overexposed 
on all images.  \\
\begin{figure}[htb]
\centerline{\includegraphics[height=50mm,width=120mm]{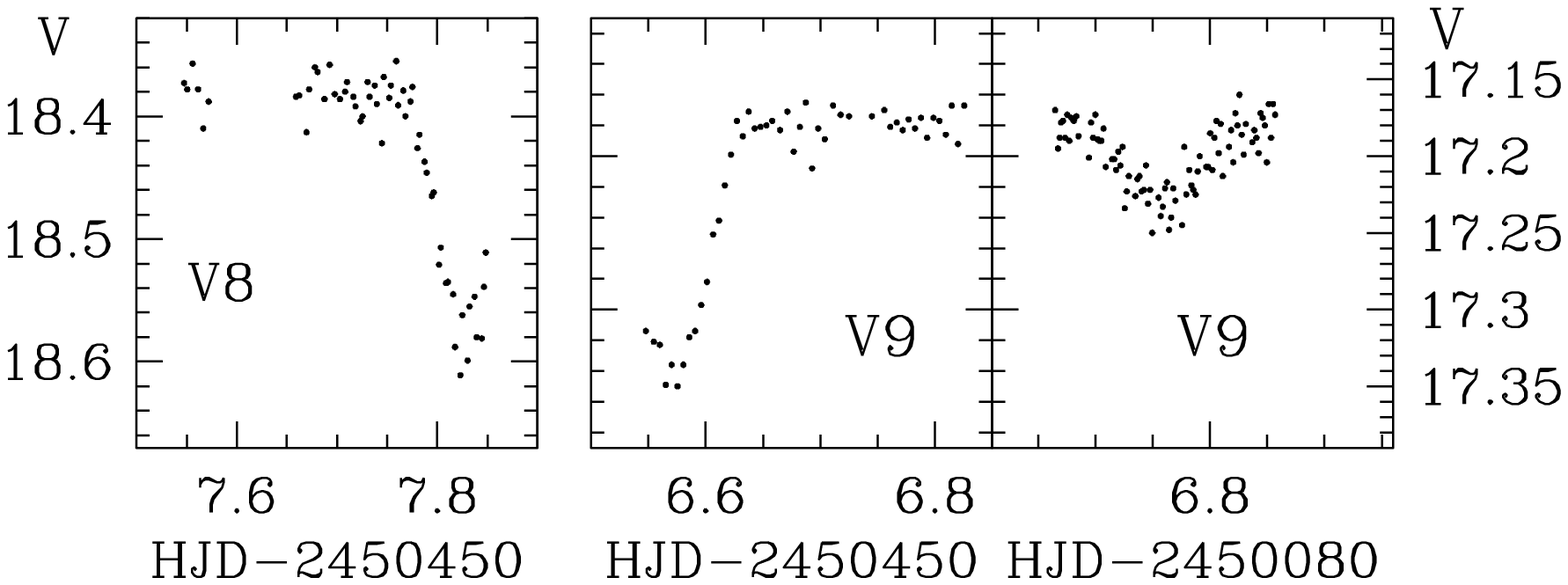}}
\FigCap{
Eclipse events in light curves of variables V8 and V9.
}
\end{figure}
%
\begin{figure}[htb]
\centerline{\includegraphics[height=50mm,width=120mm]{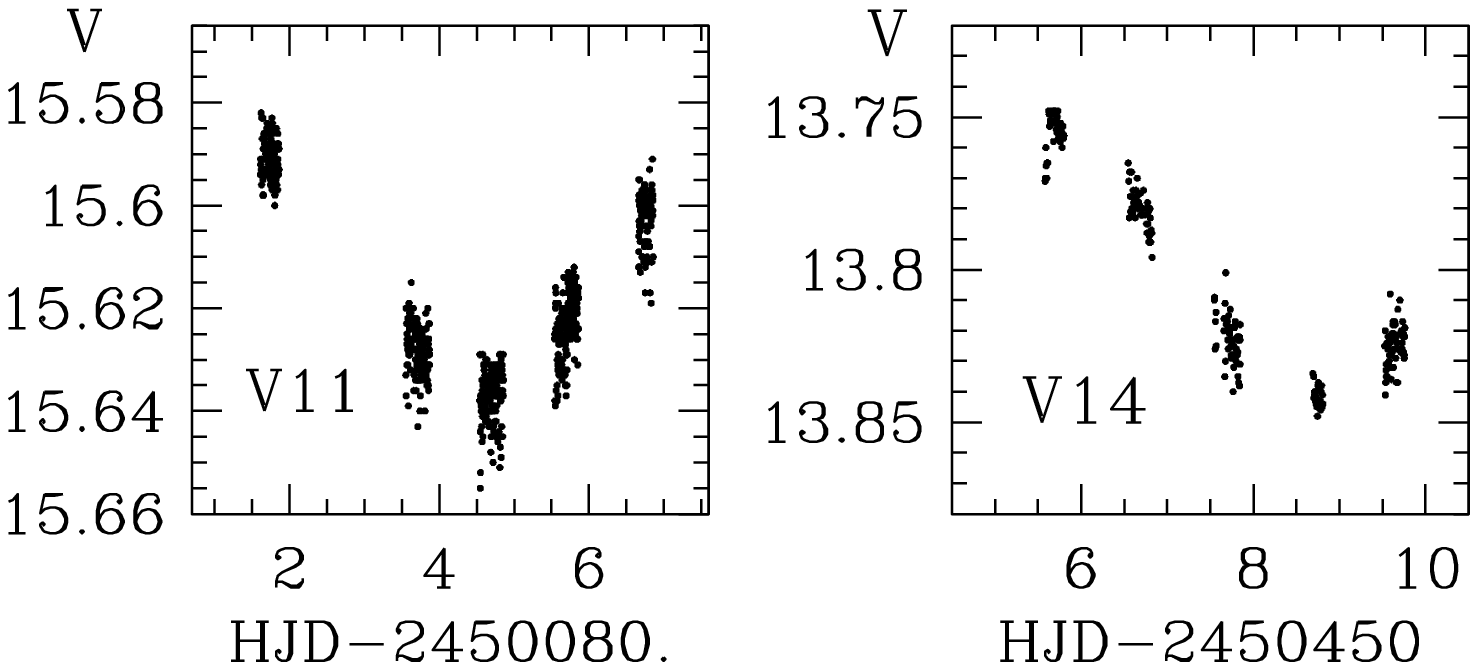}}
\FigCap{
Light curves of variables V11 and V14.
}
\end{figure}

Figure 7 shows the location of all variables from Table 2 on the
cluster color-magnitude diagram. The eclipsing binary V7 is a candidate
for a NGC~2243 blue straggler. Its light curve (see Fig. 4) is typical
for low mass-ratio semi-detached binaries. Our photometry implies
noticeable season-to-season variations of the light curve of V7. Such
variability is quite common among algols of late spectral type.
Variable V4 is the brightest eclipsing binary in the present sample.
It is located in the outskirts of the cluster. Based on the measured
radial velocity of V4 (Kaluzny et al. 2006, in preparation) one may
conclude that it is a foreground object not related to the cluster. The
cluster velocity is +55 km/sec (Friel et al 2002).  All remaining
eclipsing variables are located either on the main sequence of the
cluster or on a sequence of `photometric binaries' located slightly
above the main-sequence.  

An estimate of the absolute magnitudes of contact binaries can be made
using using the calibration due to Rucinski (2004):
\begin{eqnarray}
M_V = -4.4~ log P + 3.02~(B-V)_0 + 0.12
\end{eqnarray}
We obtain $M_{V}=4.37$, $M_{V}=3.24$ and  $M_{V}=4.09$ for V2,
V3 and V12, respectively. Assuming these absolute magnitudes, and
adopting a distance modulus for the cluster of $(m-M)_{V}=13.15$
(Anthony-Twarog et al. 2005), we derive $V_{max}=17.52$,
$V_{max}=16.39$ and $V_{max}=17.24$ for V2, V3 and V12, respectively.
These magnitudes differ by 0.14-0.34 mag from  the values listed in
Table 2.  Hence all three contact binaries are very likely  members of
NGC~2243.

A determination of the systemic radial velocities is the best way of
answering the question of cluster membership of the remaining eclipsing
variables from our list.    Finally, we note that the two reddest
identified variables --- V11 and V14 --- are candidates for cluster
subgiants. Their low amplitude variability is likely related to
so-called `spot activity' which is observed in many binaries hosting
cool sub-giants.

\begin{figure}[htb]
\centerline{\includegraphics[height=90mm,width=90mm]{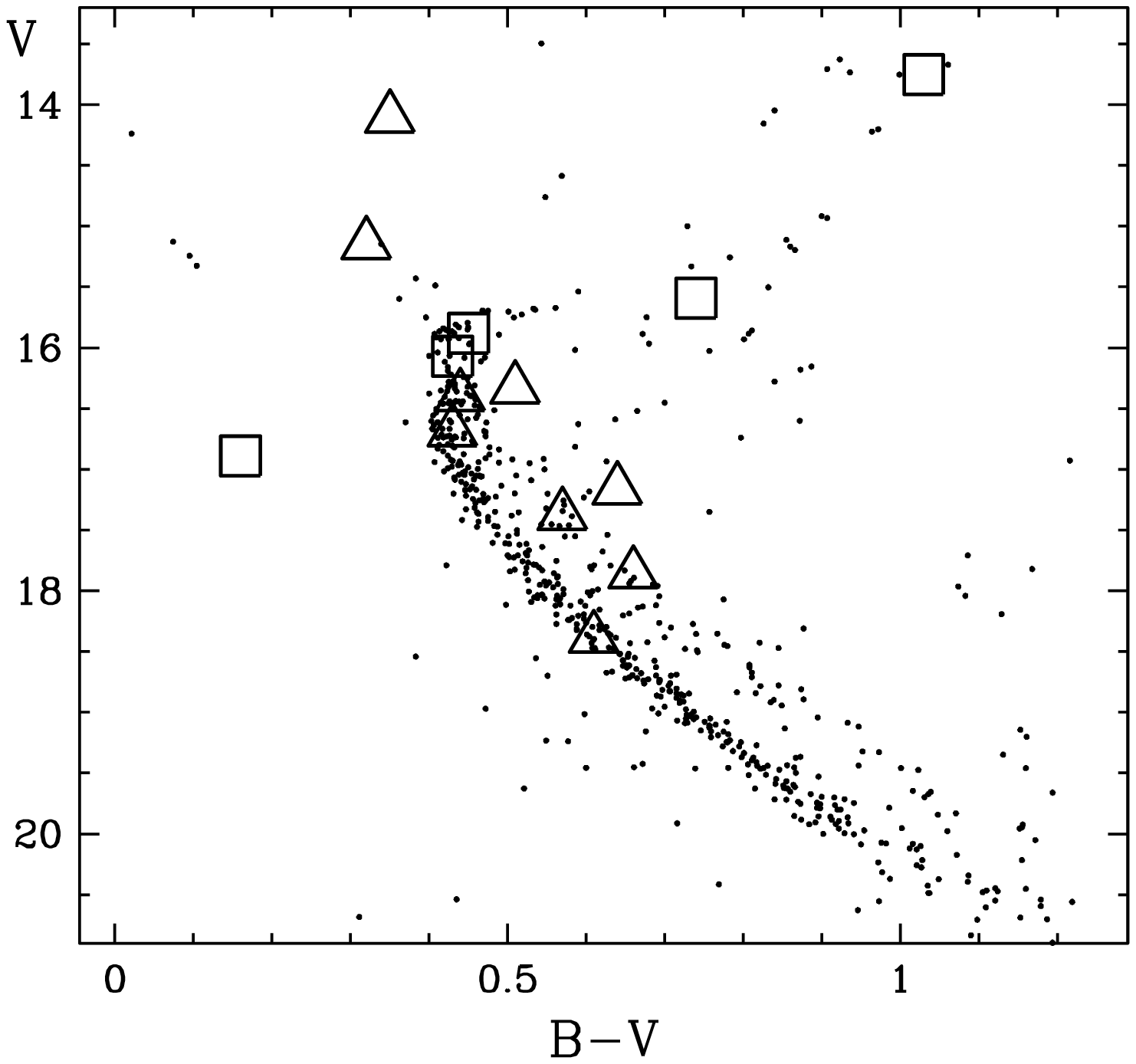}}
\FigCap{
Color magnitude diagram of NGC~2243 with positions of variables
marked: triangles -- eclipsing binaries; squares -- other objects. 
}
\end{figure}
\subsection{Period Study of  NV~CMa}

The variable NV~CMa=V1 is of particular interest, as noted in section
1.  In the second paper of this series we  plan to present a full
analysis of its photometric and spectroscopic observations, for which
the spectroscopy was collected during the last two observing seasons
(falls of 2004 and 2005).  To obtain accurate phases for the moments of
spectroscopic observations, it is necessary to have an up to date
ephemeris.  From the available data, we can derive a total of 9 times
of minimum light for NV~CMa; their values, along with errors determined
using the method of Kwee \& Van Woerden (1956), are given in Table 4.
The last of the listed minima was observed in 2005 October.  The O-C
values listed correspond to the linear ephemeris:

\begin{equation}
Min I = HJD~244 8663.70748(5) + 1.18851590(2)
\end{equation}
determined from a least squares fit to the data. The obtained fit gives
a reduced $\chi ^{2}$ equal to 1.5, which indicates that the adopted
formal errors of times of minima are slightly underestimated.  A linear
ephemeris provides a good fit and there is no evidence for any
detectable period change during the interval 1992--2005 covered by our
data.
\begin{table}
\centering
\caption{\small Times of minima and $O-C$ values determined for NV~CMa }
{\small
\begin{tabular}{lrccc}
\hline
Source & Cycle&$T_{0}$ &  Error & O-C \\
       &      & HJD-2400000  &        &     \\
\hline
KKM     & 0       &   48663.70767 &     0.00013 &   -0.00019\\
KKM     & 2.5     &   48666.67908 &     0.00020 &   -0.00031\\
run \#1 & 1195.5  &   50084.57802 &     0.00008 &    0.00022\\
run \#1 & 1196.5  &   50085.76680 &     0.00007 &   -0.00004\\
run \#2 & 1508.5  &   50456.58373 &     0.00005 &   -0.00001\\
run \#2 & 1509.5  &   50457.77201 &     0.00016 &    0.00022\\
run \#2 & 1511    &   50459.55505 &     0.00013 &   -0.00004\\
run \#4 & 4208.5  &   53665.57666 &     0.00010 &   -0.00001\\
run \#4 & 4211    &   53668.54800 &     0.00010 &   -0.00006\\
\hline
\end{tabular}}
\end{table}
\section{Discussion and Summary}

We have expanded from 6 to 14 the sample of variable stars known in the field  
of the open cluster NGC~2243. Of primary interest are 5 detached or
semi-detached binaries  which are likely members of the cluster.
Detailed analysis of these objects can provide a direct measure
of the distance to NGC~2243 based on the surface brightness method. 
Four out of five binaries are composed of stars located in different
parts of the cluster main sequence --- assuming that all 4 binaries are members 
of NGC 2243. Determination of absolute parameters for up to 8 stars
of the same age, metallicity and distance would provide a powerful test 
of evolutionary models of low mass stars.

Unfortunately, for the moment orbital periods are known with confidence
only for one detached and one semi-detached binary: V1 and V7.  We have
also obtained almost complete light curves  for these two systems. A
dedicated extended photometric survey will  be necessary to determine
periods and to obtain good coverage of eclipses for the remaining 3
detached eclipsing binaries.  In addition, spectroscopic data suitable
for accurate determination of systemic velocities of all the binaries
are needed to confidently resolve the question of their membership
status.  (It should be noted that stars in old open clusters exhibit
small dispersions of radial velocities. For example Mathieu et al
(1990) quote a value of 0.5~km/s for M67 whose age is comparable to
that of NGC 2243. Hence, measurement of radial velocities provides a
powerful discrimination between field stars and cluster members.)

There are three certain and one likely W~UMa type variables in the
central area of the cluster. The surveyed sample included about 1000
stars and is dominated by cluster members, as can be seen in Fig. 7.
Hence, the relative frequency of occurrence of contact binaries in
NGC~2243 is about 0.3-0.4\%. This is comparable to the frequency of
0.2\%  observed for binary dwarfs in the Solar neighborhood (Rucinski
2002).


\Acknow{
JK, WK \& GS were supported by grants 1~P03D~001~28 and 
76/E-60/SPB/MSN/P-03/DWM 35/2005-2007
from the Ministry of Science and Information Society Technologies, Poland.
IBT was supported by NSF grant AST-0507325.
}

\end{document}